%% file: cog2020-lowerlimb.tex
\documentclass[conference]{IEEEtran}
\usepackage{cite}

\ifCLASSINFOpdf
\else
\fi
\usepackage[pdftex]{graphicx}
\usepackage{url}
\usepackage{subfigure}
\usepackage[pdftex]{graphicx}
\DeclareGraphicsExtensions{.pdf,.jpeg,.png}

\begin{document}

\title{Lower Limb Rehabilitation in Juvenile Idiopathic Arthritis using Serious Games}
\author{Fabrizia Corona$^1$, Alex De Vita$^2$, Giovanni Filocamo$^1$, Michaela Fo\`a$^1$,\\Pier Luca Lanzi$^2$\thanks{Names appear in alphabetical order. Contact autor: pierluca.lanzi@polimi.it}, Amalia Lopopolo$^1$, and Antonella Petaccia$^1$\\
	$^1$Dipartimento della Donna, del Bambino e del Neonato\\
	Fondazione IRCCS Cà Granda Ospedale Maggiore Policlinico, Milano, Italy.\\
	$^2$Dipartimento di Elettronica, Informazione e Bioingegneria, Politecnico di Milano, Milano, Italy.
}


\maketitle

\begin{abstract}
\input{abstract}
\end{abstract}

\section{Introduction}
\input{sec_introduction}


\section{Related Work}
\label{sec:related}
\input{sec_related}
\section{Juvenile Idiopathic Arthritis}
\label{sec:juvenile_short}
\input{sec_juvenile_short}

\section{Designing Serious Games for\\JIA Lower Limb Rehabilitation}
\label{sec:design_lower_limb}
\input{sec_design_lower_limb}

\section{Experimental Results}
\label{sec:preliminary_results}
\input{sec_preliminary_results}
\section{Conclusion}
\label{sec:conclusion}
\input{sec_conclusion}


\ifCLASSOPTIONcaptionsoff
  \newpage
\fi



%


\input{cog2020-lowerlimb.bbl}

\end{document}

%% file: abstract.tex
Patients undergoing physical rehabilitation therapy must perform
	series of exercises regularly over a long period of time to improve, or at least not to worsen, their condition.
Rehabilitation can easily become  boring because of the tedious repetition of simple exercises, 
	which can also cause mild pain and discomfort.
As a consequence, 
	patients often fail to follow their rehabilitation schedule with the required regularity, thus endangering their recovery.
In the last decade, video games have become largely popular and 
	the availability of advanced input controllers has made them a viable approach
	to make physical rehabilitation more entertaining while increasing patients motivation.
In this paper, 
	we present a framework integrating serious games for the lower-limb rehabilitation
	of children suffering from Juvenile Idiopathic Arthritis (JIA).
The framework comprises games that implement parts 
	of the therapeutic protocol followed by the young patients and 
	provides modules to tune, control, record, and analyze the therapeutic sessions. 
We present the result of a preliminary validation we performed with patients at the clinic 
	under therapists supervision. 
The feedback we received has been overall very positive both from
	patients, who enjoyed performing their usual therapy using video games, and 
	therapists, who liked how the games could keep the children engaged and motivated while performing the usual therapeutic routine.

%% file: sec_introduction.tex
Physical rehabilitation therapy is a long process
	that is essential for the treatment of disabling, chronic or acute pathologies.
It involves series of tiring, and sometimes painful, exercises 
	that  patients have to perform regularly over a long period of time to improve, or not to worsen, their condition.
Unfortunately, 
	patients easily lose interest in repetitive rehabilitation exercises 
	and can become frustrated \cite{c5}, thus compromising 
	the effectiveness of therapy. 
Video games have become increasingly popular in the last decade as a mean 
	to make rehabilitation more engaging and to increase patients motivation \cite{c7}.
Initially, off-the-shelf solutions were tested in rehabilitation scenarios \cite{c10,c11,c12},
	but their application to this domain has been generally found infeasible in practice \cite{c13}.
The broad availability of advanced and inexpensive input devices capable of tracking 
	several types of body movements (e.g., 
	Leap Motion,\footnote{\url{https://en.wikipedia.org/wiki/Leap_Motion}},  
	Microsoft Kinect\footnote{\url{https://en.wikipedia.org/wiki/Kinect}}, and Intel Real Sense Depth Camera\footnote{\url{https://www.intelrealsense.com/depth-camera-d435/}}) 
	has made interacting with video games more intuitive and made them a viable approach to implement rehabilitation protocols \cite{c5}.
Accordingly, researchers and therapists developed serious games targeting specific diseases and rehabilitation protocols.

In this work, 
	we present a framework we developed for the lower-limb rehabilitation of 
	children affected by Juvenile Idiopathic Arthritis (JIA) that are following 
	the therapeutic protocol at Clinica Pediatrica G. e D. De Marchi.\footnote{\url{http://www.fondazionedemarchi.it}}
Juvenile idiopathic arthritis (JIA) is the most common type of arthritis in children under the age of sixteen.
Its cause has not been identified yet, however there are several cases of spontaneous remission.
Therapy for JIA patients aims at triggering remission while controlling pain, preserving muscle strength, motion range, and physical development \cite{Alex27}. 
Our framework integrates four video games that implement a series of rehabilitation exercises for lower-limbs
	that the young patients routinely have to perform as part of their protocol.
The framework also include 
	(i) 	a module to let the therapists parametrize the exercises for each patients,
	(ii)	a module to generate random game levels within the constraints specified by the therapists;
	(iii) 	a module to record all the actions performed by the patients in a game 
			that can be later analyzed or used to play back the session;
	(iv)	a replay module to let therapists review what a patient has done during the exercises.
All the games implement a simple adaptive mechanism
	to decrease the game difficulty (or stop the game) when the exercises are not performed within the constraints specified by the therapist.
We present the results of a preliminary validation we performed with young patients under the supervision of therapists and physicians.
Overall, we received positive feedback from the young patients,
	who enjoyed performing known rehabilitation exercises using video games,
	and therapists who were satisfied with the framework and its potentials for engaging and motivating the young patients.

The paper is organized as follows. 
In Section~\ref{sec:related}, we briefly overview the research on games for the rehabilitation of lower limbs. 
In Section~\ref{sec:juvenile_short}, we introduce Juvenile Idiopathic Arthritis, we illustrate its causes, symptoms, treatment, 
and the rehabilitation protocol followed at Clinica Pediatrica G. e D. De Marchi (where our study was conducted).
In Section~\ref{sec:design_lower_limb}, we present our framework and discuss
the requirements we received from the therapists, the system structure, some of its functionalities, 
and the rehabilitation games it provides.
In Section~\ref{sec:preliminary_results}, we describe the experimental setup and discuss 
some results from a preliminary validation we performed with patients at the clinic under the supervision of therapists.
Finally, in Section~\ref{sec:conclusion}, we draw some conclusions and delineate future research directions.

%% file: sec_related.tex
We present a brief overview of the research on games for the rehabilitation of lower limbs. We refer the interested reader to one of the available surveys on video games for rehabilitation 
\cite{lohse,ONeil2014,Bonnechere2018,DBLP:journals/tciaig/PirovanoMBLB16,Christensen2016,Grampurohit2019,Tay2018} for a broader view of the area.

Rehabilitation of lower limbs using serious games has been less explored compared to upper limbs \cite{lanzi:2018:rehabilitation,DBLP:journals/sensors/BorgheseEMMPP19}. 
This probably because, in the past,
	input devices for hands and arms were more available or easier to develop.
However, the introduction of inexpensive input devices like the Kinect and Wii Fit, 
	made it possible to capture all body movements without wearing constraints
	and enabled the development of rehabilitation games targeting lower limbs.
There are very few works targeting arthritis patients accordingly we take a broader view considering also other pathologies that share many design challenges with our work.
Imam et al. \cite{Imam2018ACS} performed an online survey targeting physical and occupational therapists 
	to collect data about the use of commercial games for lower limb prosthetic rehabilitation in Canada. 
The data they collected from 82 therapists showed that 46.3\% used commercial games and 
	that almost all of them used the Wii Fit.
The perceived benefits included improved weight shifting and balance, 
	increased motivation when complementing traditional therapy; 
in contrast, 
	the most reported perceived barriers/challenges
	were due to the  lack of time and familiarity with the games.

Lai at al. \cite{Lai_Warburton_2017} integrated upper and lower limb rehabilitation 
	using virtual reality exergaming to improve the functional status of children with Duchenne muscular dystrophy.
Carlos Luque-Moreno et al. \cite{cul17} developed a systematic review of the literature to describe the different virtual reality and interactive videogames solutions applied to the lower extremity of stroke patients. 
Jaffe et al. \cite{cul18} compared two training groups one doing rehabilitation 
	with real obstacles and one using virtual ones. 
In the second group, a head-mounted device was used to observe the simultaneous registration of the legs’ real movement, introducing virtual stationary images of obstacles and getting a patient’s feedback. The virtual obstacle training generated greater improvements in gait velocity compared with real training during the fast walk test and the self-selected walk test \cite{cul17}.
Gil-Gómez et al. \cite{cul19} compared an intervention program with the Nintendo Wii Balance Board (WBB) with eBaViR to a conventional physiotherapy treatment in patients with brain damage. 
Patients using the Wii Balance Board had a significant improvement in static balance, compared to patients who go through traditional therapy. With regard to dynamic balance, there were no differences between study groups [17].

Fritz et al. \cite{cul20} compared an experimental group that used Nintendo Wii (Wii Sports and Wii Fit) and Play Station (Eye Toy Play 2 and Kinetic) with a control group that underwent no intervention. No statistically significant differences in the comparison between or within groups were found, either in the short term or in the follow-up process \cite{cul17}.
Chen et al. \cite{cul21} used a human-computer interactive videogame based rehabilitation device (LLPR) for the training of lower limbs muscle power in the elderly. Twenty over forty participants in the exercise group received a 30-min training, twice a week, using the LLPR system. The LLPR system allowed participants to perform fast speed sit-to-stand movements. Twenty age-matched participants in the control group performed slow speed sit-to-stand movements, as well as strengthening and balance exercises, with the same frequency and duration. Results showed that in the exercise group, all the mechanical and time parameters demonstrated significant improvement. In control group, only the maximal vertical ground reaction force improved significantly. For clinical assessments (balance, mobility, and self-confidence), exercise group showed significantly better scores.

%% file: sec_juvenile_short.tex
Juvenile Idiopathic Arthritis (JIA), formerly known as juvenile rheumatoid arthritis (JRA),
	is a broad term that describes a clinically heterogeneous group of arthritis of unknown cause \cite{c19}.
The term identifies several disease categories,
	each of which has distinct methods of presentation, 
	clinical signs, symptoms and, in some cases, genetic background.
%
Research suggests that JIA is an autoimmune disease. 
In autoimmune diseases, white blood cells cannot tell the difference between
	body's own healthy cells and germs like bacteria and viruses. 
Thus, the immune system, which is supposed to protect the body from these harmful invaders, 
releases chemicals that can damage healthy tissues and cause inflammation and pain \cite{Alex23}. 
It is unclear what causes the illness, 
	it seems related both to genetic and environmental factors, 
	which result in the heterogeneity of the illness \cite{Ravelli2007767}.
About one child in every one thousand develops some type of juvenile arthritis \cite{Alex4}.

\subsection{Treatment}
A cure for juvenile idiopathic arthritis is yet to be found, but there are many cases of spontaneous remission. 
	Therapy aims at inducing such remission while controlling pain and preserving range of motion, muscle strength, physical and psychological development \cite{Alex27}. 
Children with chronic arthritis are treated with a combination of four main factors: 
	(i) the pharmacological management of the disease, 
	(ii) physical therapy, 
	(iii) nutrition, and 
	(iv) orthopedic surgery. 
Pharmacological treatment typically begins as soon as the disease is discovered since the sooner it starts, the less probable it is that there will be permanent aftereffects \cite{Alex28}. 
Physical therapy is used to minimize pain, maintain and restore functionality, prevent deformity and disability, correct wrong compensatory behavior. 
Nutrition targets long-term management and typically involves nutritional and vitamin supplementation. 
Orthopedic surgery has a limited role in management of chronic arthritis in young children while it might be used in the older children for joint contractures, dislocations, or joint replacement. 

\subsection{Physical Therapy}
Joint inflammation is one of the main symptoms of juvenile idiopathic arthritis that, if left untreated treated, can result in the loss of articular functionality and in the consequential worsening of the patient’s quality of life. Moreover, children tend to interiorize incorrect compensatory postures or movements that persist even after a full recovery. 
These compensations burden on muscles and other joints, leading to new possible physical problems. Accordingly, physical therapy focuses on helping patients managing their symptoms and improving their self-sufficiency rather than healing the inflammation. Therapists guide children in the process of understanding their moving capabilities so as to reduce their fear of pain and the families propensity to overprotect. 
Physical therapy has to be customized for each patient and usually take place both at the hospital (or at another designated structure) and at home. 
%
%


\subsection{Lower Limb Rehabilitation}
Rehabilitation of the lower limbs for juvenile idiopathic arthritis targets both large joints (e.g., knees) and small joints (e.g., ankles).
In this work, we focused on the rehabilitation protocol for small joints as carried out at the Clinica Pediatrica G. e D. De Marchi
	that comprises five exercises asking patients to maintain their balance on uneven surfaces while wearing only socks.
In the first exercise (Figure \ref{fig:alex33}), the patient rests in the center of a special foam rubber pillow. 
To maintain its balance, the body has to compensate the waning of the feet in the pillow. All the lower limbs system is involved (hips, knees, ankles and muscles).
In the second exercise (Figure \ref{fig:alex34}), the roll pillow stays on the ground while the patient sits on a chair. The pillow is made of the same foam rubber of the square pillow of the previous exercise. The patient places her feet on the pillow and rolls it back and forward. In the advanced version of the same exercise, the patient increases the pressing force, and extends the roll angle of the ankles. This exercise is also useful to stimulate the tactile perception of the lower limbs. In the third exercise (Figure \ref{fig:alex35}), the patient places the feet on the squared balance board. In the easy version, the patient sits on a chair, like the previous exercise, and swings the board back and forward working on knees and ankles; in the advanced version, the patient stands up in equilibrium and use all the lower limbs system (hips, knees, ankles and muscles). An even more difficult version, which is the same used by football players, uses a circular balance board with 360 degrees of freedom (Figure \ref{fig:alex36}). In the fourth exercise (Figure \ref{fig:alex37}), the patient stands on the aero-step. The aero-step consists in a couple of connected air chambers: if one chamber is pressed down, the air flows to the other chamber raising it up. The patient shifts the body weight from one leg to the other, using all the lower system (hips, knees, ankles and muscles) to win the air resistance of a chamber and finding an overall balance. In the fifth exercise (Figure \ref{fig:alex38}), the patient walks a route of pillows and other obstacles. The pillows have similar shapes but different consistence. Some are light and soft, others are heavy and tough, some allows the feet to sink, and others support the body weight. The obstacles are plastic surfaces with different shapes and different slopes. The patient walks back and forth along the route, adapting to the different surfaces. Every few time, the therapist changes the order of the pillows without being seen by the patient. This way, the patient doesn't learn the schema, and has to constantly work on adapting and balancing. It is a difficult but very important exercise that puts the whole lower limbs system to work (hips, knees, ankles and muscles). It is also very important at cognitive level since, due to the change in pillow configuration, the patient has to adapt in real time to different surfaces, just like it would happen in everyday life. 
Accordingly, this exercise also helps boosting the patient confidence and reduces the risk of injury in the normal shifts. 
In this work, we focused on this exercise due to the importance it has in the rehabilitation protocol. 

\input{fig_lower_limbs}

%% file: fig_lower_limbs.tex
\begin{figure*}[h]
	\subfigure[square pillow]{
	\label{fig:alex33}
	\includegraphics[width=.30\textwidth]{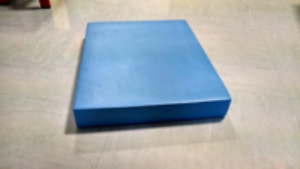}
	}%
	\qquad	
	\subfigure[roll pillow]{
	\label{fig:alex34}
	\includegraphics[width=.30\textwidth]{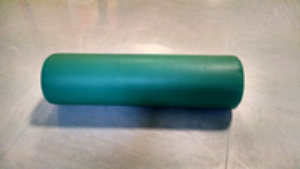}
	}%
	\qquad
    \subfigure[]{%
	    \includegraphics[width=.30\textwidth]{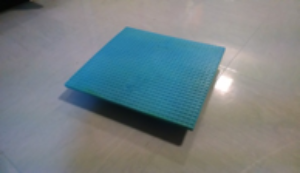}
        \label{fig:alex35}
    }\\
    \qquad
    \subfigure[thumb flexion]{%
	    \includegraphics[width=.30\textwidth]{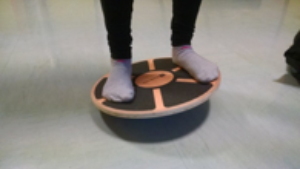}
        \label{fig:alex36}
    }%
    \qquad
    \subfigure[thumb abduction]{%
	    \includegraphics[width=.30\textwidth]{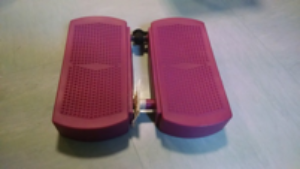}
        \label{fig:alex37}
    }%
    \qquad
    \subfigure[thumb opposition]{%
	    \includegraphics[width=.30\textwidth]{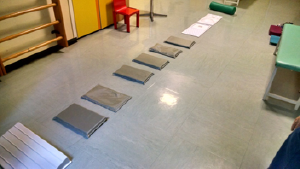}
        \label{fig:alex38}
    }
\caption{Exercises for lower limb rehabilitation: \subref{fig:alex35} squared balance board; \subref{fig:alex36} circular balance board; \subref{fig:alex37} aero step;
\subref{fig:alex38} pillow path.}
\end{figure*}

%% file: sec_design_lower_limb.tex
%
We developed the rehabilitation games based on the therapeutic protocol followed at Clinica Pediatrica G. e D. De Marchi. 
The games are part of a framework that implements the requirements and desiderata we received from the therapists 
as well as the most important features that the literature in this area suggests \cite{c13,c14,c31,c32}.

\subsection{Requirements and Desiderata}
Initially, we organized several meetings with the therapists 
	who showed us the rehabilitation protocol and demonstrated several exercises with patients. 
We agreed to focus on the exercises involving a route of pillows of different consistence (Figure~\ref{fig:alex38})
	and to use Microsoft Kinect V2 for tracking patients movements.
Therapists asked us
	to have the possibility to tune each game for a specific patient,
	to record the data about the patients posture and movements during the games, and 
	to receive a quantitative feedback about how long and how frequently patients played the games.

\subsection{Rehabilitation Framework Architecture}
The rehabilitation games are part of an integrated framework  that provides 
	(i) support for patient profile management, 
	(ii) game/therapy customization, 
	(iii) patient monitoring, and (iv) data recording.
A patient is associated to a profile containing the information about her therapy, the games she played, and statistics about each session. 
The profile is anonymous and no personal information is stored. 
Patients are identified with nicknames and 
	only therapists know who is the patient associated to a nickname. 
For the sake of simplicity, 
	each game has a set of default parameters defined by the therapists that can be easily customized for each profile. 
Therapists can manage the creation of the profiles and, for each one of them, they can view the replays of the sessions and visualize 
	the several trends associated with each patient.


\subsection{Design of Rehabilitation Games}
We needed our games to be as entertaining as possible while keeping their therapeutic value. 
We opted for an intuitive mechanics and gameplay that children could quickly learn to play without long training sessions. 
We designed a system that rewards only correct actions (therapeutic movements) and
	does not penalize possible errors: 
	score is incremented when players do something good and remains the same when mistakes are made. 
Thus, children that have more problems and are prone to do more errors, 
	will not see their score dramatically decrease for their errors---which would also be likely to cause a loss of motivation. 

Juvenile idiopathic arthritis affects both male and female children with a considerable age gap.
Accordingly, we created two themes around the same game mechanics in which set our games to appeal a range of patients as broad as possible: 
	(i) a dungeon crawler theme, in which the player is a mage that faces monsters and traps;
	(ii) a garden theme, in which the player is a bee that has to collect pollen from flowers.
We increased variety by giving the possibility to change several parameters such as
	the game length, number of lives, time allowed to move from one pillow to another, or any combination of them.
%

\begin{figure}
	\begin{tabular}{cc}
		\parbox{.48\columnwidth}{
			\includegraphics[width=.48\columnwidth]{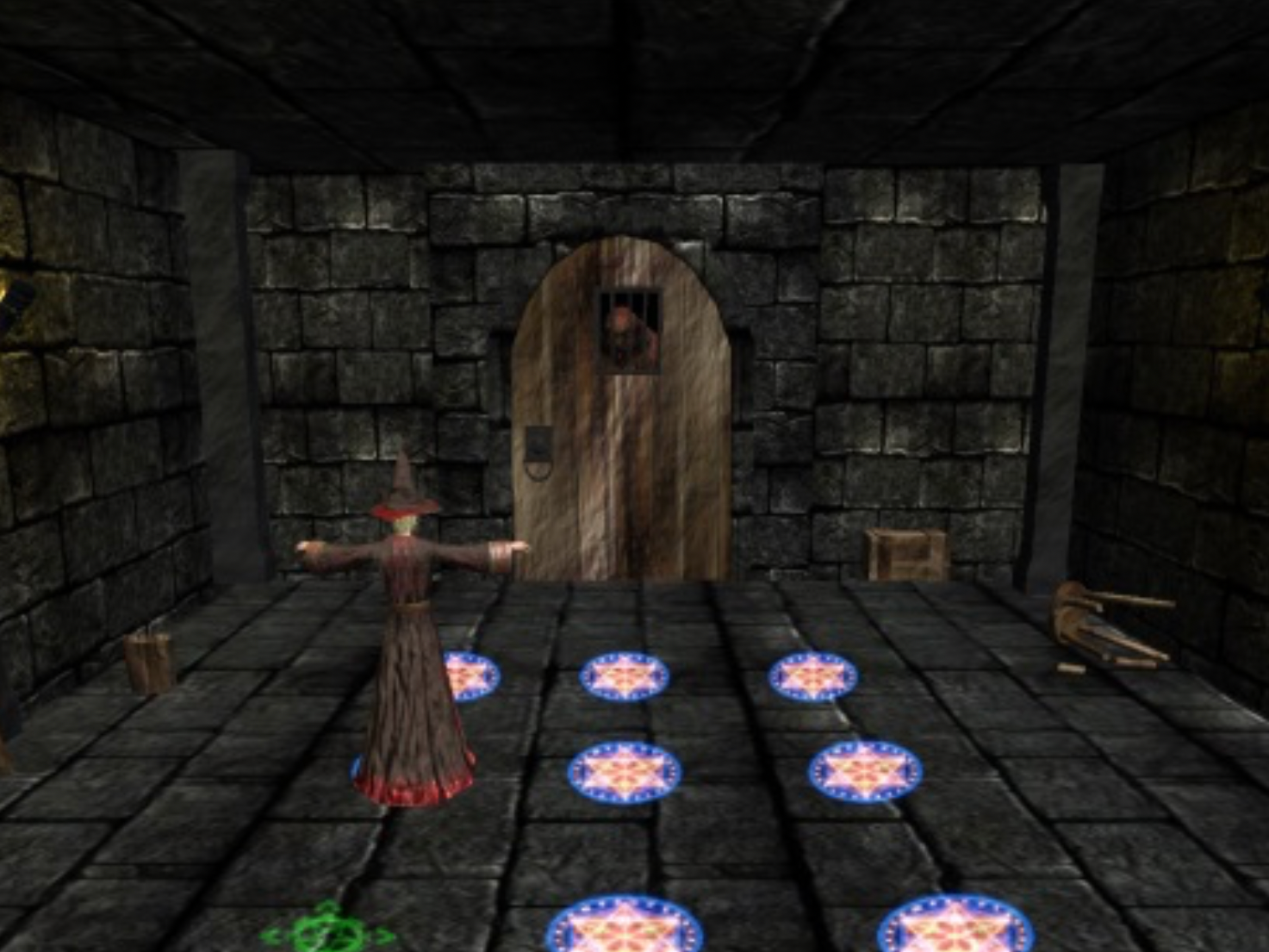}
		} & 
		\parbox{.48\columnwidth}{
			\includegraphics[width=.48\columnwidth]{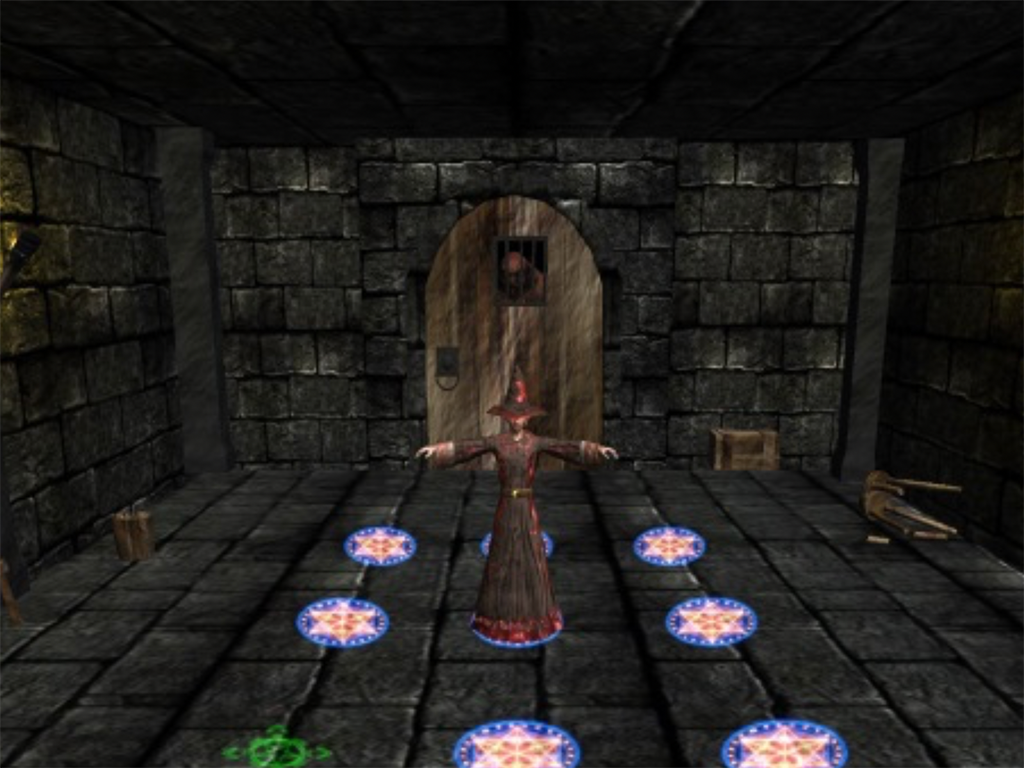}
		} \\
		(a) & (b) \\
	\end{tabular}
	\caption{Player camera position (a) third-person view; (b) front-facing, mirrored, view.}
	\label{fig:player_camera}
\end{figure}

We selected two game mechanics that could support the rehabilitation exercises with the pillows and possibly make them fun. 
One is an endless runner that uses three pillows positioned on a horizontal line. The player controls her avatar (either a mage or a bee) that has to dodge continually incoming obstacles or reach flowers by changing her position. This mechanics implements a rather simple exergame since children need only to move horizontally and thus can easily maintain their spatial reference. The second mechanics uses nine pillows positioned over a 3x3 grid and it resembles the mechanics of well known dance games in which players have to take a certain position in a certain amount of time. In our case, we have a countdown and children have to move to a highlighted position to increase their score. The position is either a safe position in the mage dungeon scenario or a flower full of pollen in the bee scenario. This type of game is challenging also at cognitive level since the avatar can be either in the typical third person view  (Figure~\ref{fig:player_camera}(a)), or in a front facing view, thus mirroring the player (Figure~\ref{fig:player_camera}(b)). Accordingly,  players need to think how to move in order to end up on the right pillow depending on the avatar orientation. The therapist and the patient can select what view they prefer before the game begins.

\subsection{Data Recording and Replay}
Rehabilitation games allow to collect the game’s data in order to analyze the patient’s performance. Using these data, it is possible to keep track of the patient’s progress, and also to have a quantitative and qualitative feedback about her movement proficiency. In our framework, games save both 
(i) the 3D spatial coordinates of every skeleton’s joints of the player, and 
(ii) the video captured by the camera. The former is useful to see if the therapy is helping the patient in reducing her movement limitations and in correcting the bad compensatory postures. 
The latter is useful to see what patients did during the games.

The replay mode allows therapists to see how the patient performed the exercise. This feature is useful both because it provides the chance to see exercises guided by a colleague, without the therapist being there; and also because, overlaying data about the posture to the video provides a qualitative feedback 
that is more intuitive and immediate than the usual data plots. This helps therapists to punctually find possible mistakes in the exercises and to show them to the patient, in order to let her understand what she did wrong.

\begin{figure}
	\includegraphics[width=\columnwidth]{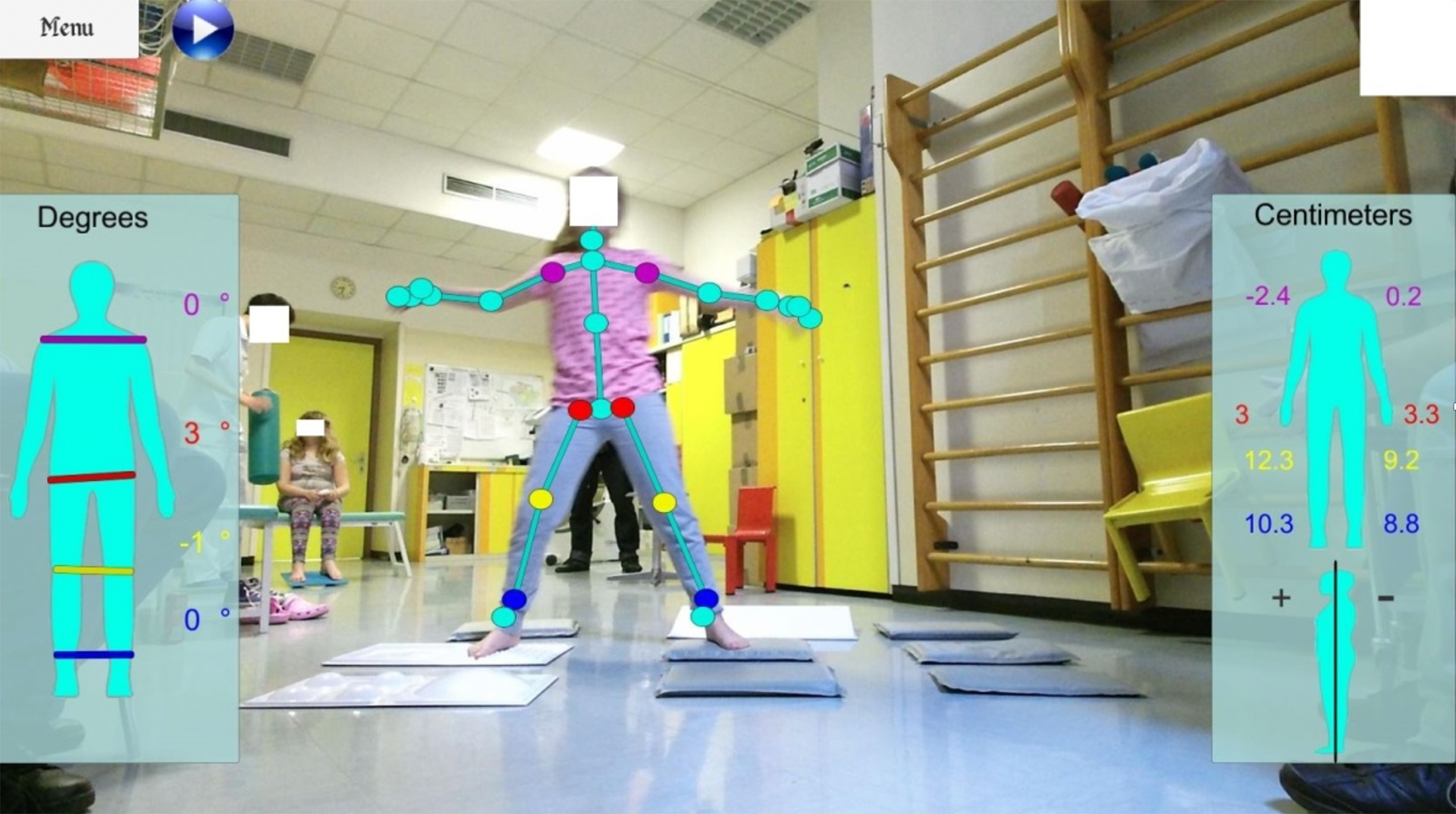}
	\caption{The replay view.}
	\label{fig:replay}
\end{figure}

Figure~\ref{fig:replay} show the replay view with the skeleton data overlaid on the captured video.
On the left of the screen, we show the posture degrees of shoulders, hips, knees, ankles. This helps to identify wrong postures interiorized by the child to compensate the affected mobility. On the right side of the screen we estimate, based on the Kinect data, the depth distances of the joints of interest from the spine base, which is the skeleton’s fulcrum. Even if we focused on lower limbs, the replay view also include data about the shoulders because they are a good marker to detect wrong postural behaviors.

\subsection{Games Setup}
The games require a setup phase to create the virtual playground by positioning the pillows on the floor in front of the Kinect sensor.
This phase might be repeated for each patient and 
	the configuration may be modified several times during each session.
Accordingly, we designed setup to be as simple as possible. 
At first, the therapists place the pillows on the floor arranged 
either into a horizontal line of three pillows
or into a three by three grid of nine pillows. 
Then,
the therapist activates the automatic setup and asks the patient to step in three different pillow positions.
The Kinect sensor uses the three positions to estimate all the remaining pillow positions so that the whole setup takes a couple of minutes.
Note that, three is the minimum number of spots to create an accurate 3x3 grid, or a line of three pillows. 
At each position, the patient raise her hand over the head and waits for the confirmation feedback from the screen. There is also a confirmation button that the therapist can press to confirm the patient position in case of physical impediments. 
Since the Kinect v2 cannot recognize generic objects, we could not use it to automatically recognize the pillows. We proposed to use specially marked pillows but the therapists considered it unacceptable as they wished to have the freedom to use any support available. 
Figure~\ref{fig:setup_ll_4} shows the grid creation and game setup phase; 
Figure~\ref{fig:setup_ll_1} shows an exercise grid made of pillows and sheets of paper used only for the patients’ spatial reference; Figure~\ref{fig:setup_ll_2}, shows a more difficult configuration using pillows and obstacles; Figure~\ref{fig:setup_ll_3} shows a grid with an additional carpet underneath it, used 
to reduce the pillow shifting and to modify the overall pillow pressure resistance.

\input{fig_ll_setup}

\input{fig_games}

\subsection{Games Using a Three-Pillows Line}
\input{ssec_1x3}

\subsection{Games Using a 3x3 Pillow Grid}
\input{ssec_3x3}

%% file: fig_ll_setup.tex
\begin{figure*}
\centering
    \subfigure[]{%
	\includegraphics[width=.23\textwidth]{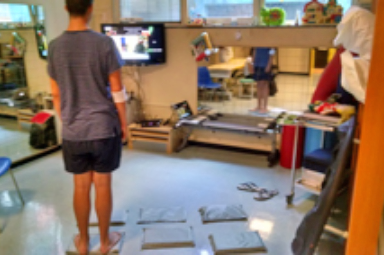}
	\label{fig:setup_ll_4}
	} %
	~
    \subfigure[]{%
	    \includegraphics[width=.23\textwidth]{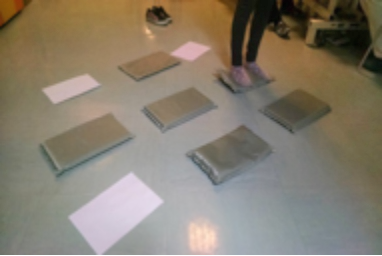}
        \label{fig:setup_ll_1}
    }%
    ~
    \subfigure[]{%
	    \includegraphics[width=.23\textwidth]{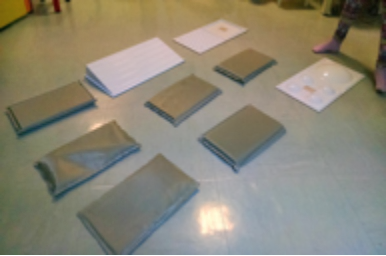}
        \label{fig:setup_ll_2}
    }%
    ~
    \subfigure[]{%
	    \includegraphics[width=.23\textwidth]{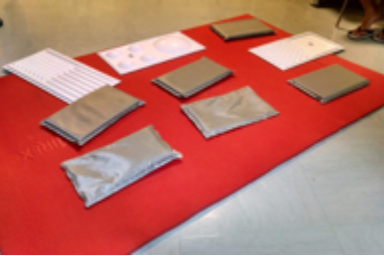}
        \label{fig:setup_ll_3}
    }

\caption{Setup for the lower-limb rehabilitation games: 
	\subref{fig:setup_ll_4} grid creation and game setup;
	\subref{fig:setup_ll_1} grid with pillows and paper;
	\subref{fig:setup_ll_2} grid with pillows and obstacles;
	\subref{fig:setup_ll_3} grid with pillows, obstacles and carpet.
}
\end{figure*}

%% file: fig_games.tex
\begin{figure*}[h]
	\subfigure[]{
	\label{fig:1x3_mage}
	\includegraphics[width=.23\textwidth]{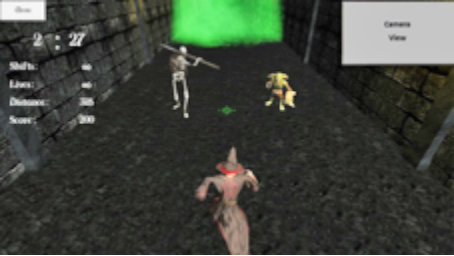}
	}%
	~
	\subfigure[]{
	\label{fig:1x3_bee}
	\includegraphics[width=.23\textwidth]{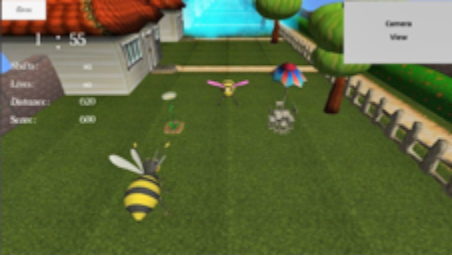}
	}%
	~
    \subfigure[]{%
	    \includegraphics[width=.23\textwidth]{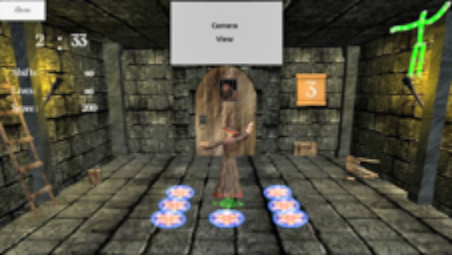}
        \label{fig:3x3_mage}
    }%
	~
    \subfigure[]{%
	    \includegraphics[width=.23\textwidth]{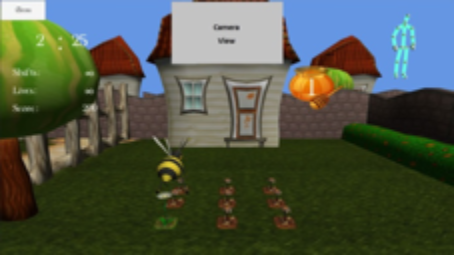}
        \label{fig:3x3_bee}
    }
\caption{Video games for lower limb rehabilitation using three pillows positioned on a horizontal line 
	(\subref{fig:1x3_mage} Devix the mage and \subref{fig:1x3_bee} Beatrice the bee) and 
	nine pillows on a 3x3 grid (\subref{fig:3x3_mage} Devix the mage and \subref{fig:3x3_bee} Beatrice the bee).}
\end{figure*}


%% file: ssec_1x3.tex
This set of games targets lateral movements. Each game uses three pillows positioned on one horizontal row and
  the player has to move horizontally from one pillow to another while keeping her balance.
In the first game (Figure~\ref{fig:1x3_mage}), ``Devix the mage and the endless run," the player’s is a mage who has to run avoiding obstacles and monsters.
The player's avatar (i.e., the mage) is at the bottom of the screen and continuously moves forwards. 
The green marker in front of the avatar indicates the safe pillow (so in this case the safe position is in the center) while the monsters indicates the bad ones.
Every time monsters appear, the player score increases;
when the player’s avatar hits an obstacle, the mage says ``Ouch" while the obstacle cracks a sound based on its nature. 
Obstacles and monsters are randomly generated and their frequency is controlled by the therapist who can setup all the game parameters (length, speed, obstacles/monsters frequency).
The second game targets a younger audience by using a cartoon graphics while maintaining the same game mechanics (Figure~\ref{fig:1x3_bee}). In \textit{Beatrice the bee and the flower run}, the player is a bee that has to collect flowers avoiding other bees and obstacles. As before, the bee is at the bottom of the screen and flies continuously forwards. The flower on the ground indicates the safe pillow (so in this case the safe position is in the left) while the obstacles indicates the bad ones. Every time obstacles or bees spawn, the score increases.


%% file: ssec_3x3.tex
This set of games targets both lateral, forward and backward movements. Each game uses nine pillows 
  positioned on a 3x3 grid. In ``Devix the mage and the magical traps," 
  	the player must avoid magical traps, identified by stars on the floor (Figure~\ref{fig:3x3_mage}),
  	and reach safe spots, identified by green markers.
%
%
 The score is positioned on the top left of the screen while on the right of the screen, an ancient scroll shows the countdown to the next traps activation. When the countdown reaches zero, the traps are triggered and the game checks the patient’s position. If the patient is on the safe position (i.e. the correct pillow), the score increases and sound effects celebrate the mage; if the patient is on the wrong pillow,  the mage is hit by magical flames, no points are gained or lost and the mage plays a funny ``Ouch" sound. Every player’s movement is mapped over the avatar, so she can have fun shaking his arms and legs around.
In ``Beatrice the bee and the honey preparation,"
  the player is a bee that has to collect all the healthy flowers to prepare the honey (Figure~\ref{fig:3x3_bee}). Dried flowers on the ground identify bad positions that should be avoided, while healthy flowers are the safe spots. Again, a flower corresponds to a physical pillow placed on the floor. The score is on the top left while a honey jar on the top right shows the countdown to the next position change. When this countdown reaches zero, if the player is on the right pillow, therefore the bee is on the healthy flower, the score increases and the game produces a happy sound; if the player is on the wrong pillow, (thus the bee is on the dried flowers), no points are not gained nor lost, the bee produces a funny noise. Due to the differences in the structure between a human body and a bee, only the translations are mapped on the player’s avatar. The bee has however a default animation.
Both games have a mirrored version in which the avatar is facing the player and the obstacles appear from the bottom of the screen.
When the rehabilitation session starts, the therapist decides which version of the game to play based on protocol and the player preferences. 
Typically, therapists tend to use the plain version for easy rehabilitation sessions
	and the mirrored version to enrich the rehabilitation with a cognitive component. 

%

%% file: sec_preliminary_results.tex
\begin{figure}
\includegraphics[width=\columnwidth]{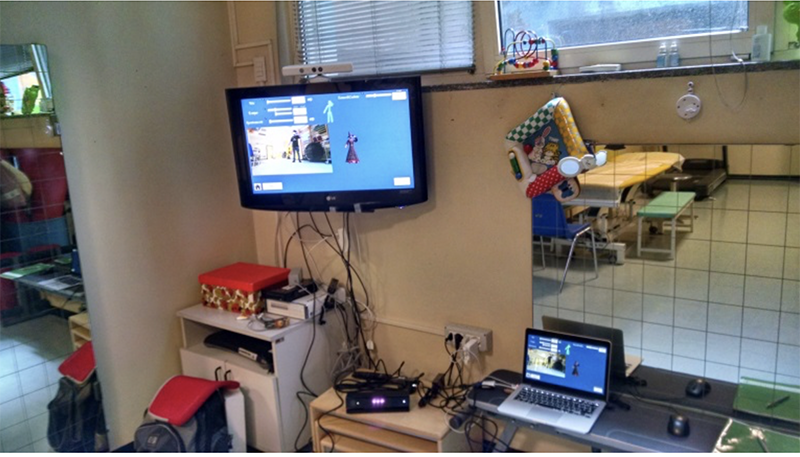}
\caption{Experimental setup at the clinic therapy gym.}
\label{fig:gym}
\end{figure}

We performed a series of experiments to validate our framework 
	with young patients under the supervision of two therapists and one physician in the therapy gym at the clinic;
	the Kinect sensor was placed under the monitor and connected to a notebook (Figure~\ref{fig:gym}).
	
\subsection{Setup and Initial Tuning}
We performed a series of sessions to tune the games and also to collect initial feedback from patients and the therapists
	both from a rehabilitation viewpoint and in terms of players' engagement.
The sessions involved two females subjects and one male subject who played early versions of ``Devix the mage and the magical traps'' on  a 3x3 grid and ``Devix the mage and the endless run"
	on a three pillows line. Patients enjoyed the games and appreciated the new way of doing their rehabilitation routine. They also provided some important feedback to improve the games. For example, in this early stage, the virtual playground had no marker on the safe pillow and highlighted only the positions to be avoided. Patients suggested we should also highlight the safe pillows to help their movements. 
Therapists' feedback was also very positive as they noticed that patients concentrated on the score, paid less attention to their impediments, and could move faster than usual.	
Interestingly, therapists initially set to 15 seconds the time available to patients to shift from one pillow to another; this value was set based on the traditional rehabilitation sessions they performed with these patients. However, after the tuning, we had to reduce the time to ten seconds because the patients, while playing, were able to move faster than in usual rehabilitation and thus asked us to reduce the time constraint.

\subsection{Therapeutic Sessions}
Next, we performed two sessions with patients followed at the clinic.
The sessions involved three females (F4, F5, F6) 
	in the first session and one male and three females (F3, F6, M7, F8) in the second session.
F4 was ten years old; F5 was twelve years old and F6 was nine; F3 was fifteen years old and was also one of the participants 
	to the tuning session; M7 was fifteen years old; 
	F4 had knees problems; F8 had ankle problems.
All the sessions took place in the clinic gym and the setup was the same used for tuning. 
%
%
The feedback was again very positive. F3 enjoyed the sessions more than the initial tuning. F6 showed some lack of initiative during the grid creation phase, probably due to her young age, but then she demonstrated great effort and agility during the games she played. 
F4 really enjoyed the rehabilitation session despite she had knee problems and also some weight issue that made exercising mode strenuous. F5 was initially detached but her attitude improved during the session. F8 gave good feedback despite her ankles problems.

\begin{figure}[t]
	\centering
	\includegraphics[width=.8\columnwidth]{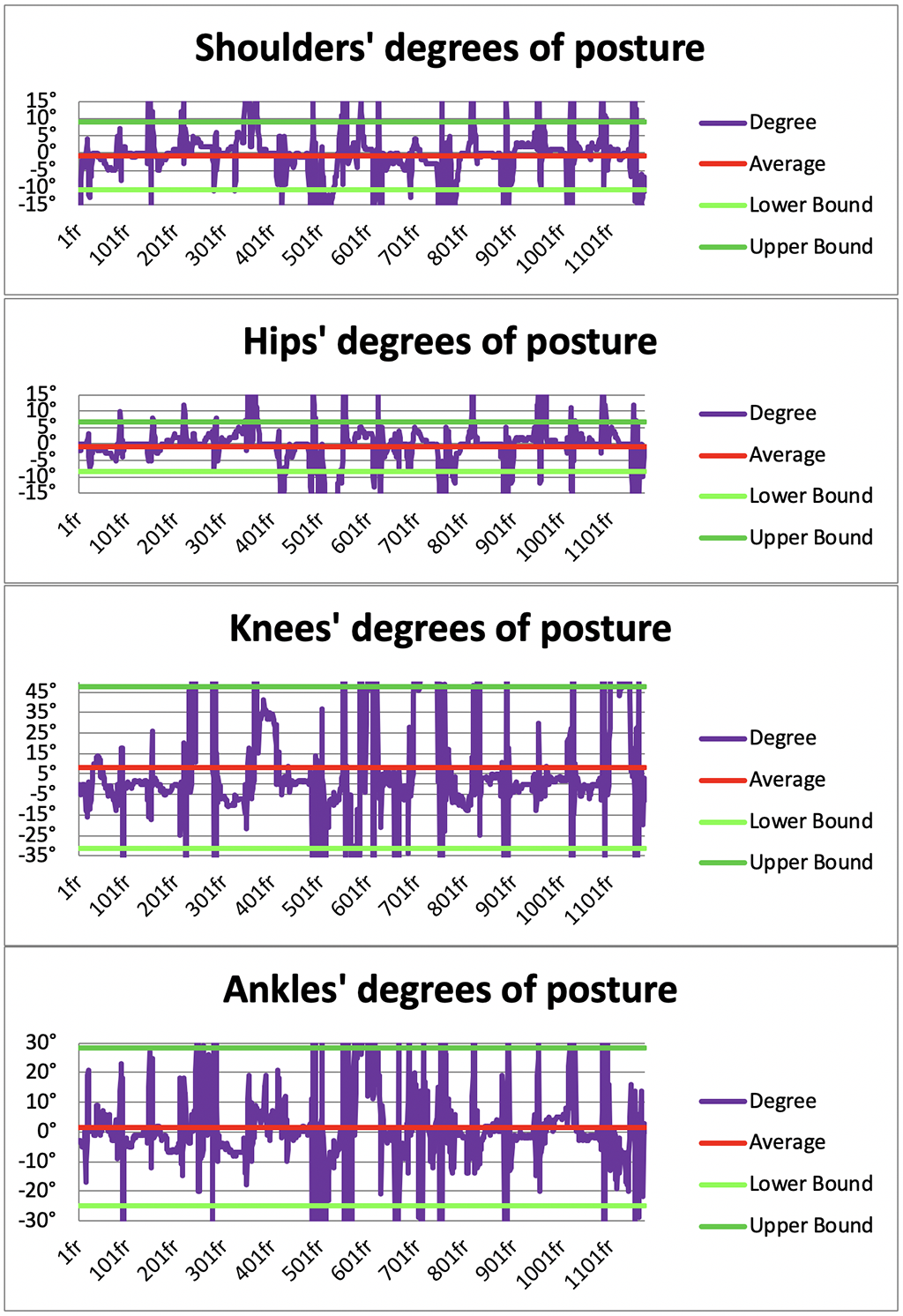}
	\caption{Performance F6, first session on a 3x3 grid.}
	\label{fig:f6_3x3s1}
\end{figure}

As an example, we discuss the results of two rehabilitation sessions of F6. 
In the first  session, F6 played ``Beatrice the bee and the flowers run'' using a three pillows line;
	``Devix the mage and the magical traps" on a 3x3 grid;
		``Devix the mage and the endless run"  using a three pillows line.
In the first game, F6 demonstrated good posture and a tendency to keep the feet together.
Figure~\ref{fig:f6_3x3s1} shows the data collected during the second game on the 3x3 grid.
The plots reveal a worsening of the knees posture. Actually, F6 kept a better lower limbs positioning, with the legs more spaced to be more agile during the shifts. During the resting phases, however, she shifted the body weight over one leg, resulting in one lower knee.
In the third game,  F6 reached her best performance with high peaks during the shifting between pillows due to her considerable agility and speed. Her movement were cleaner and the resting position more balanced.


During the second therapeutic session, F6 played the same sequence of games.
In the first and third game, F6 again demonstrated good postural behaviors with high peaks during the shifting phases due to her high agility. The data showed an improvement of the legs positions, which were more spaced and balanced, and in her speed and strength.

%% file: sec_conclusion.tex
We designed a framework integrating serious games 
	to support the physical therapy of children affected by Juvenile Idiopathic Arthritis to the lower limbs. 
We worked with therapists and patients to create games that could be both medical relevant and fun to play. 
During the experimental sessions, we received very positive feedback both from the patients and the therapists involved.
Young patients enjoyed the games and also helped us by suggesting features to improve their design. 
Therapists noted that patients were more focused on the score than their impediments.
Accordingly, they moved faster than usual and were able to perform movements that, in a traditional rehabilitation session, 
	they would perform only if been forced to. 
Therapists found the replay feature very useful for the evaluation of the therapy.
Kinect was usually very effective for tracking skeleton data accurately. 
In some extreme scenarios, with the most severe JIA cases, we experienced some noise and variability in the data collected for the lower limbs. 
However, the replay allowed therapists to promptly and easily recognize the issue and continue the evaluation process without problems.

In the future, we plan to add more games to target a broader range of exercises such as the ones for the rehabilitation of single joints
	which we could not implement due to the limitation of Kinect tracking.
Accordingly, we also plan to test other technologies such as Azure Kinect\footnote{\url{azure.com/Kinect}} 
	and Intel Real Sense Depth Camera\footnote{\url{https://www.intelrealsense.com/depth-camera-d435/}} 
	although the higher price might be an issue for families and hospitals.	 
Therapists suggested that the mental challenge introduced by the mirrored/not mirrored versions of the games, 
	could be useful in the rehabilitation of  cerebellitis.
Thus, we are considering to extend the applications of our framework to include such types of pathologies.